\documentclass[%
reprint,
superscriptaddress,
amsmath,amssymb,
aps,
]{revtex4-1}

\usepackage{graphicx}
\usepackage{dcolumn}
\usepackage{bm}
\usepackage{amsmath,amssymb,amsfonts}
\newcommand{\new}[1]{{\color{black}{#1}}}
\usepackage{xcolor}
\usepackage[normalem]{ulem}
\usepackage{soul}

\usepackage[pagebackref=false]{hyperref}

\begin{document}
	
	\author{L.~Banszerus}
	\email{luca.banszerus@rwth-aachen.de.}
	\affiliation{JARA-FIT and 2nd Institute of Physics, RWTH Aachen University, 52074 Aachen, Germany,~EU}%
	\affiliation{Peter Gr\"unberg Institute  (PGI-9), Forschungszentrum J\"ulich, 52425 J\"ulich,~Germany,~EU}
	\author{T.~Fabian}
	\affiliation{Institute for Theoretical Physics, TU Wien, 1040 Vienna, Austria, EU}
	\author{S.~M\"oller}
	\author{E.~Icking}
	\affiliation{JARA-FIT and 2nd Institute of Physics, RWTH Aachen University, 52074 Aachen, Germany,~EU}%
	\affiliation{Peter Gr\"unberg Institute  (PGI-9), Forschungszentrum J\"ulich, 52425 J\"ulich,~Germany,~EU}
	\author{H.~Heiming}
	\affiliation{JARA-FIT and 2nd Institute of Physics, RWTH Aachen University, 52074 Aachen, Germany,~EU}%
	\author{S.~Trellenkamp}
	\author{F.~Lentz}
	\affiliation{Helmholtz Nano Facility, Forschungszentrum J\"ulich, 52425 J\"ulich,~Germany,~EU}
	
	\author{D.~Neumaier}
	\affiliation{AMO GmbH, Gesellschaft f\"ur Angewandte Mikro- und Optoelektronik, 52074 Aachen, Germany, EU}
	\affiliation{University of Wuppertal, 42285 Wuppertal, Germany, EU}
	\author{M.~Otto}
	\affiliation{AMO GmbH, Gesellschaft f\"ur Angewandte Mikro- und Optoelektronik, 52074 Aachen, Germany, EU}
	\author{K.~Watanabe}
	\affiliation{Research Center for Functional Materials, 
		National Institute for Materials Science, 1-1 Namiki, Tsukuba 305-0044, Japan}
	\author{T.~Taniguchi}
	\affiliation{International Center for Materials Nanoarchitectonics, 
		National Institute for Materials Science,  1-1 Namiki, Tsukuba 305-0044, Japan}%
	\author{F.~Libisch}
	\affiliation{Institute for Theoretical Physics, TU Wien, 1040 Vienna, Austria, EU}
	\author{C.~Volk}
	\author{C.~Stampfer}
	\affiliation{JARA-FIT and 2nd Institute of Physics, RWTH Aachen University, 52074 Aachen, Germany,~EU}%
	\affiliation{Peter Gr\"unberg Institute  (PGI-9), Forschungszentrum J\"ulich, 52425 J\"ulich,~Germany,~EU}%

\title{Electrostatic detection of Shubnikov-de-Haas oscillations in bilayer graphene by Coulomb resonances in gate-defined quantum dots}%
	
	
	\date{\today}
	
\begin{abstract}
A gate-defined quantum dot in bilayer graphene is utilized as a sensitive \new{probe} for the charge density of its environment. Under the influence of a perpendicular magnetic field, the charge carrier density of the channel region next to the quantum dot oscillates due to the formation of Landau levels. This is experimentally observed as oscillations in the gate-voltage positions of the Coulomb resonances of the nearby quantum dot. From the frequency of the oscillations, we extract the charge carrier density in the channel and from the amplitude the shift of the quantum dot potential. We compare these experimental results with an electrostatic simulation of the device and find good agreement.
\end{abstract}
\maketitle   

\section{Introduction}
Sharp Coulomb resonances in quantum dots (QDs) and single electron transistors (SETs) can be used to sense changes in the electrostatic potential defining the 
corresponding charge island. Thus, these systems are sensitive probes of changes of their
electrostatic environment with remarkable high precision~\cite{Shapir2019May}. SETs are for example used to read out changes of charge states in neighboring QDs~\cite{Elzerman2003Apr,Barthel2010Apr} or they can be mounted on a piezo scanner to perform scanning SET microscopy~\cite{Martin2007Nov}. 
In this work, we use a gate defined QD in bilayer graphene (BLG) to detect changes in the density of states (DOS) of the doped BLG channel region next to the QD acting as lead. When applying a perpendicular magnetic field, we observe oscillations in the potential of the QD. These are caused by charge carrier density fluctuations in the lead regions due to the formation of Landau levels and the tuning of their energies 
as function of the applied perpendicular magnetic field~\cite{Zhang2005Nov}. We determine the average charge carrier density in the narrow leads from the frequency of the Shubnikov-de-Haas oscillations and the shift of the QD potential from their amplitude. 
These results are compared with a full electrostatic simulation of the device and good agreement is found. 

\begin{figure}[ht!]
\centering
\includegraphics[draft=false,keepaspectratio=true,clip,width=\linewidth]{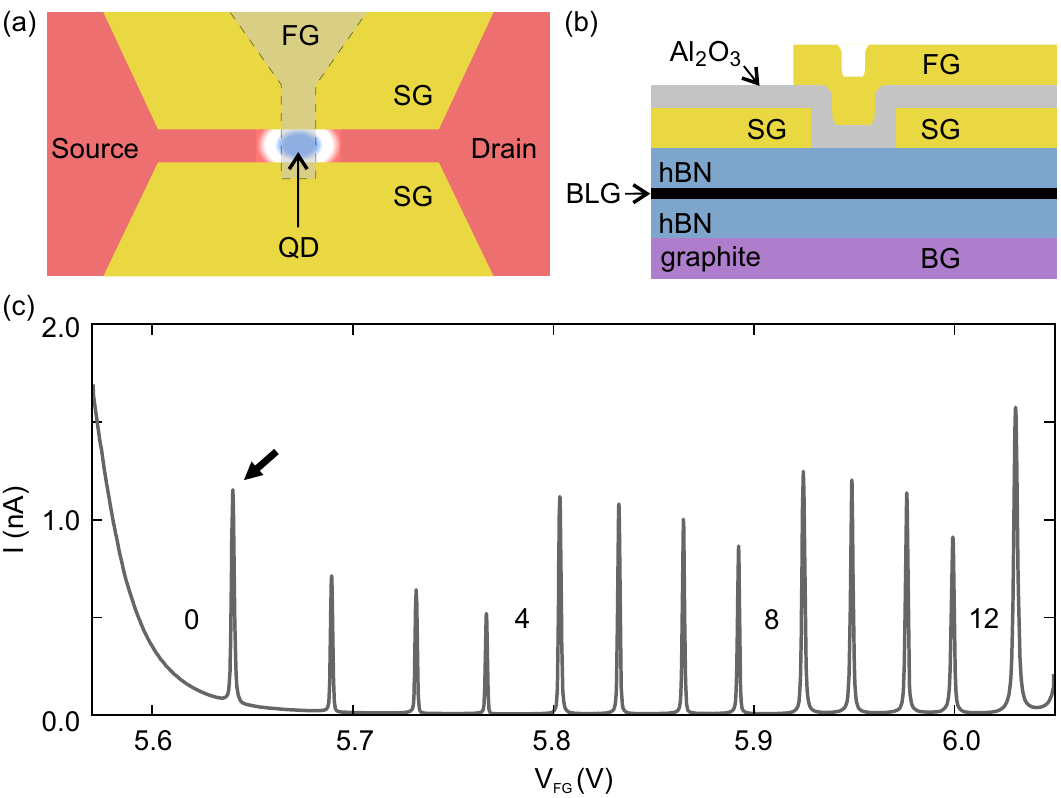}
\caption[Fig01]{
\textbf{(a)} Schematic illustration of the device: A narrow, highly doped channel is formed between the two split gates (SGs), connecting source (S) and drain (D) contacts. Using a gold finger gate (FG), a quantum dot is defined by forming a small n-doped island in the p-doped channel. 
\textbf{(b)} Schematic cross section  through the sample. The BLG is encapsulated between two flakes of hBN and placed on a graphite gate. On top, gold split and finger gates are deposited and separated by a layer of Al$_2$O$_3$. 
\textbf{(c)} Source-drain current through the device as function of $V_{\mathrm{FG}}$ at a constant bias voltage of $V_{\mathrm{b}}=200$~$\mu$V and an electron temperature below 100~mK.
} 
\label{f1}
\end{figure}

\section{Fabrication}
The device studied in this work consists of a BLG flake encapsulated between two crystals of hexagonal boron nitride (hBN), placed on a graphite gate using the conventional van-der-Waals stacking technology~\cite{Wang2013Nov}. Similar to previous work studying BLG gate-defined quantum point contacts~\cite{Overweg2018Jan,Kraft2018Dec,Banszerus2020May} and quantum dots~\cite{Eich2018Aug,Eich2018Jul,Banszerus2018Aug,Banszerus2020Mar,Kurzmann2019Aug,Kurzmann2019Jul}, two layers of gold gates are evaporated on top: A pair of split gates (SGs) is used to form a 150~nm wide conducting channel connecting the source and drain reservoirs of the device (see Fig. 1(a)). On top, separated by a 30~nm thick film of atomic layer deposited Al$_2$O$_3$, we place a gold finger gate (FG) with a width of 70~nm (see Figs. 1(a) and 1(b)).

\section{Device Characterization}
All measurements are performed in a ${}^3\mathrm{He}/{}^4\mathrm{He}$ dilution refrigerator at a base temperature of 15~mK. Applying $V_{\mathrm{BG}}=-4.5~V$ to the graphite back gate and $V_{\mathrm{SG}}=3.4~V$ to both split gates, we open a band gap in the BLG underneath the SGs, leaving only a narrow conductive channel between the SGs~\cite{Oos2007Dec,Zhang2009Jun,McCann2006Oct}. 
Similar to recent works, we can form a QD underneath the FG by locally overcompensating the applied back gate voltage~\cite{Eich2018Aug,Eich2018Jul,Banszerus2020May,Banszerus2020Mar,Kurzmann2019Aug}. A small n-doped island is created underneath the FG, which is separated from \new{the p-doped} channel by the band gap acting as a tunnel barrier (see schematic in Fig.~1(a))~\cite{Banszerus2020Mar,Eich2018Jul}. 
We measure the current through the device as a function of $V_{\mathrm{FG}}$ (Fig.~1(c)). At $V_{\mathrm{FG}} \ll 5.6~$V the entire channel is p-doped and well conductive. Increasing $V_{\mathrm{FG}}$ leads to a decrease of the current as the Fermi level crosses the band gap underneath the FG. Sharp Coulomb resonances appear above $V_{\mathrm{FG}} \approx 5.6$~V. As observed in previous work, the resonances are grouped in quadruplets, representing the spin and valley degeneracy of BLG~\cite{Eich2018Jul,Eich2018Aug,Banszerus2020Mar,Knothe2020Jun}. From finite bias spectroscopy measurements (not shown), we extract a charging energy of $E_c \approx 6$~meV, a total capacitance of the QD of $C_\mathrm{tot}\approx27$~aF and a finger gate capacitance of $C_\mathrm{FG}\approx3.3$~aF. Describing the QD using a model of a disk shaped plate capacitor, we extract a QD diameter of $d\approx70$~nm, which is in reasonable agreement with the lithographical dimensions of the gate electrodes.

\begin{figure}[]
\centering
\includegraphics[draft=false,keepaspectratio=true,clip,width=0.95\linewidth]{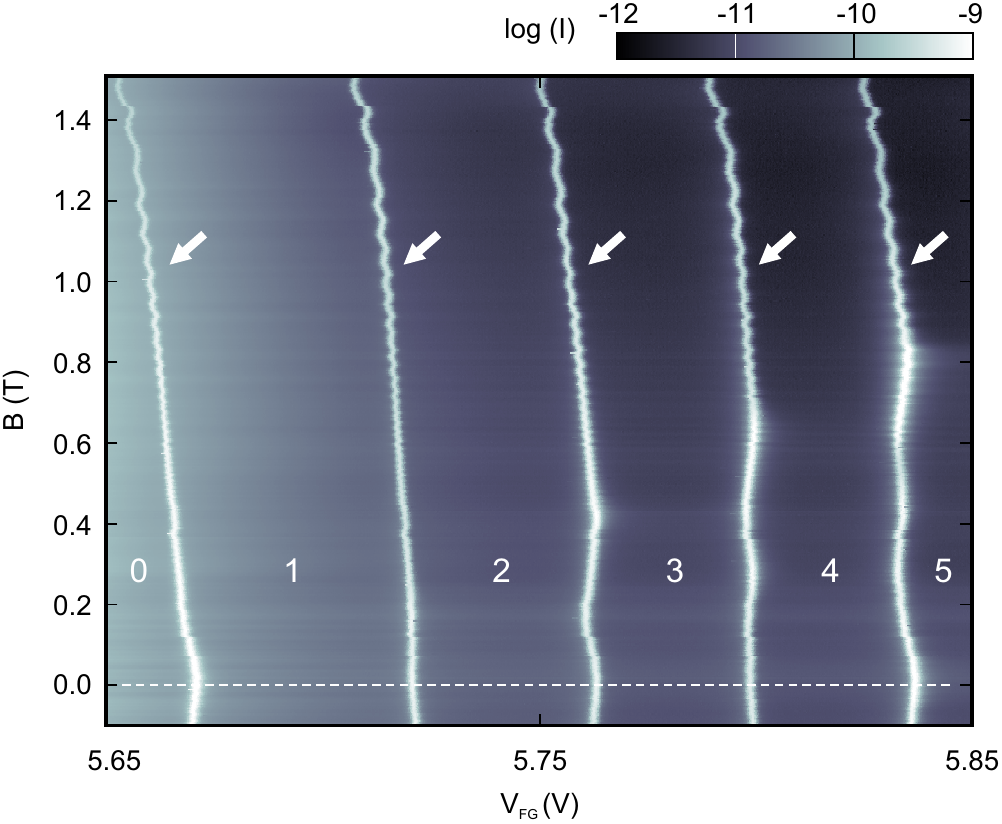}
\caption[Fig02]{
Source-drain current through the QD in the regime of the first five Coulomb peaks (\new{The arrows indicate the Coulomb resonances and the }dot occupation is labeled by white numbers) as a function of $V_{\mathrm{FG}}$ and perpendicular magnetic field, $B$, at a constant bias voltage of $V_{\mathrm{b}}=200\mu$V.
} 
\label{f2}
\end{figure}

\begin{figure*}[]
\centering
\includegraphics[draft=false,keepaspectratio=true,clip,width=\linewidth]{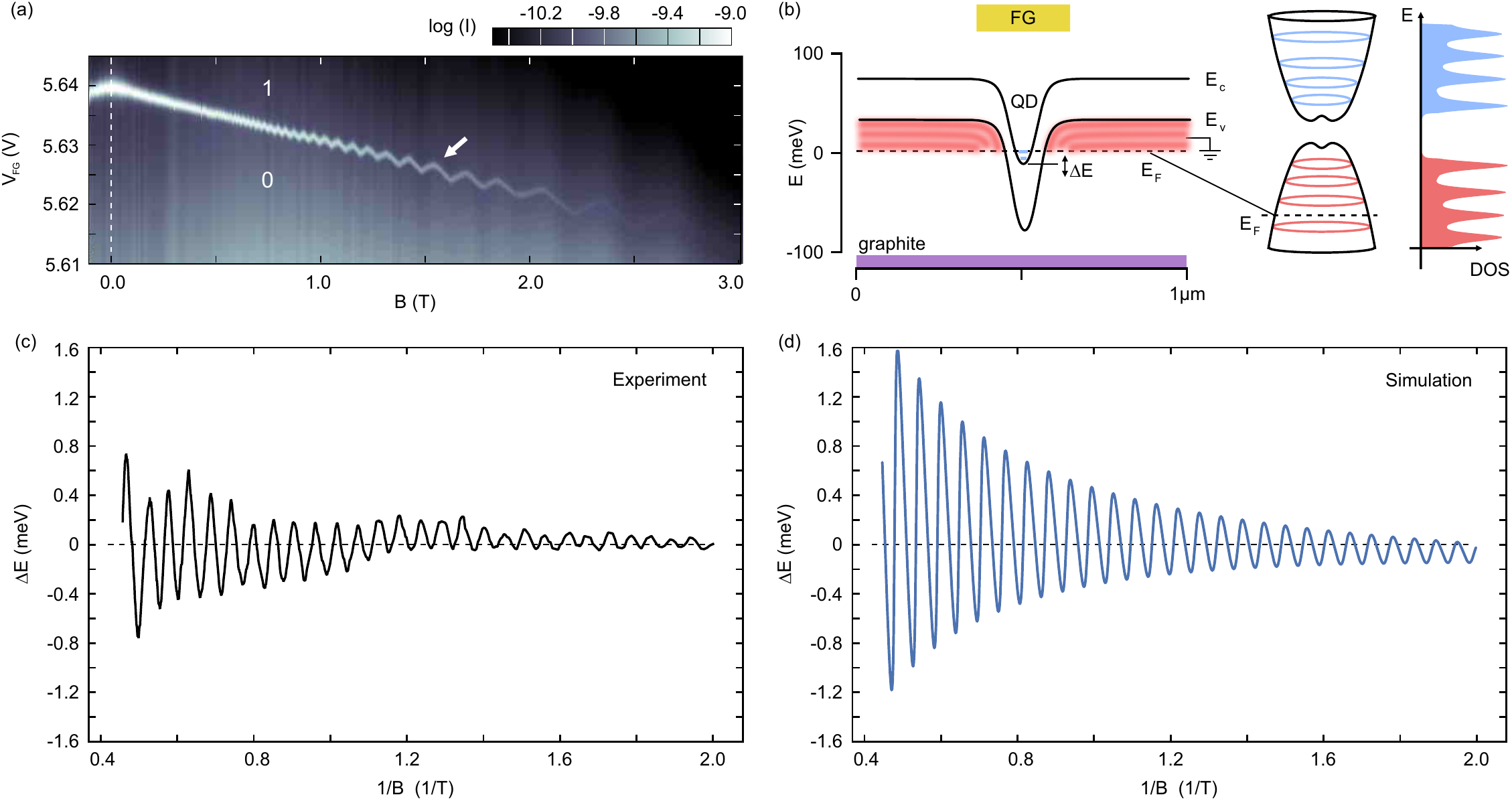}
\caption[Fig03]{
\textbf{(a)} Logarithmic current through the device around the transition of $N=0$ and $N=1$ electrons in the dot as function of the applied $V_{\mathrm{FG}}$ and the perpendicular magnetic field, $B$, at a fixed bias of $V_{\mathrm{b}}=200$~$\mu$V. \textbf{(b)} Electrostatic calculation of the band alignment around the QD. The FG induces a p-n-p junction where an electron QD is formed. The density of states in the channel regions is modulated as function of a perpendicular magnetic field due to the formation of Landau levels. \textbf{(c)} Extracted shift of the QD potential as function of the inverse perpendicular magnetic field. \textbf{(d)} Simulated shift of the QD potential (minimum in b) as function of the inverse perpendicular magnetic field.}
\label{f3}
\end{figure*}

Sharp features in the gate characteristic, such as Coulomb resonances, are well suited to monitor small changes in the electrostatic environment of the QD, which result in shifts of the Coulomb peaks~\cite{Guttinger2008Nov}. In order to demonstrate that the QD is a suitable \new{probe} for small changes in the charge carrier density $n$ in the channel and lead region, we apply a perpendicular magnetic field, $B$, to the device, resulting in Shubnikov-de-Haas (SdH) oscillations in the channel. The modulation of the DOS with $B$ results in a change of the quantum capacitance between the channel and the back gate, causing charge to enter or exit the channel region through the source and drain reservoirs.
Tuning the magnetic field has the advantage that we are able to change $n$ without changing any gate voltages applied to the device, which would result in a shift of the QD potential itself.
Fig.~2 shows magnetotransport data as a function of the gate voltage $V_\mathrm{FG}$ and $B$. 
The presented measurement shows the first five Coulomb resonances of the QD filling up the dot with $N=1,2,3$ and $4$ electrons (see labels in Figs.~1(c) and 2). 
\new{The effect of small magnetic fields below 0.5~T is to shift levels with valley-dependent linear slopes by coupling to the topological orbital magnetic moment of bilayer graphene ~\cite{Eich2018Jul,Thess1996Jul,Tans1997Apr}.
The alternating slopes of higher occupation numbers are due to multiple quantum mechanical level crossings of states of the two valleys as function of $B$.}
Thus, we focus on the first Coulomb resonance, i.e. the transition from $N=0$ to $N=1$ electron in the QD (see arrow in Fig.~1(c)), where $N$ is the occupation number of the dot. The single particle spectrum of BLG QDs is well understood~\new{\cite{Eich2018Jul,Kurzmann2019Jul,Pereira2007Sep,daCosta2016Feb}, where} the energy of the ground state of $N=1$ \new{just} decreases linearly in $B$. Since the DOS of the QD depends only slightly on $B$, we neglect quantum capacitance effects for the capacitances between the source/drain lead and the QD. 

Fig.~3(a) shows a (rotated) close-up of the first Coulomb resonance. The \new{linear} decrease in \new{$V_\mathrm{FG}$} with increasing $B$-field results from the valley magnetic moment in BLG, which couples to the magnetic field. A similar effect is also known from carbon nanotubes as described in Refs.~\cite{Thess1996Jul,Tans1997Apr,Eich2018Jul}. For magnetic fields exceeding $B~\approx 0.5$~T, the position of the Coulomb resonance starts to oscillate at a frequency decreasing with magnetic field. 

\new{In contrast to the linear shifts discussed above, where the slope is alternating for different Coulomb resonances, the oscillatory behaviour is a common energy shift of all Coulomb resonances (see Fig.~2), indicating that this is an external, electrostatic effect. 
The perpendicular magnetic field gives rise to Shubnikov-de-Haas oscillations of the density of states in the p-doped channel region (see Fig.~3(b)). This results in an oscillating charge carrier density in the channel region (see Fig.~3(b)) and effectively gates the QD.
Thus, besides the direct influence of the gates, the Coulomb peak position also depends on the magnetic field: First, due to valley-dependent topological orbital magnetic moment, and second, due to the $1/B$ oscillation of charge carrier density in the lead.}
Note that the development of clean Landau levels/gaps and well separated edge states is suppressed in the $W = 150~$nm wide channel region, as the cyclotron orbit $r_\mathrm{c}= m^* v_\mathrm{F} /(eB)$ is larger than half the channel width for $B < 3$~T (here $m^*=0.033m_e$ is the effective mass of BLG where $m_e$ is the electron mass, $v_\mathrm{F}$ is the Fermi velocity and $e$ is the elementary charge). Instead, we observe merely a moderate modulation of the density of states, which is also reflected in the rather sinusoidal oscillations of the Coulomb peak position indicating that no clean Landau gaps are formed yet. 

We determine the gate-voltage position of the Coulomb peak from Fig.~3(a) and subtract a linear fit to account for the valley Zeeman effect. 
The lever arm $C_\mathrm{FG} / C_\mathrm{tot}$ allows to convert the shift in gate voltage $\Delta V_{\mathrm{FG}}$ into a shift of the QD potential $\Delta E$.
The QD potential oscillates periodically as function of the inverse magnetic field $1/B$ (see Fig.~3(c)), as expected for SdH oscillations. The amplitude increases with increasing magnetic field (decreasing $1/B$), as the cyclotron radius decreases and gets closer to $W/2$, resulting in \new{a stronger modulation of the density of states}~\cite{Terres2016May}. 
The charge carrier density in the channel determines the period of the oscillations given by~\cite{Novoselov2006Feb} 
\begin{equation}
    \Delta \left( \frac{1}{B} \right)=\frac{4e}{hn}.
    \label{e:Oscillation}
\end{equation} 
\new{From the periodicity of the SdH oscillations we extract a charge carrier density of} $n=1.72\times10^{12}$~cm$^{-2}$.

\section{Simulation}
\new{Our goal is to simulate the response of the system to a magnetic field, given the measured charge carrier density.}
 We model the experiment in a two-step process.
 First, we solve the \new{Poisson equation} at $B=0$~T. In a second step, we add a magnetic field by making the total charge carrier density $B$-field dependent. In a self-consistent way we solve the Poisson equation for a two-dimensional cut along the transport direction (see red line Fig.~4), we iteratively add charge to the BLG sheet until its Fermi energy equals the electrostatic potential, which we implemented with fenics~\cite{AlnaesBlechta2015a,LoggWells2010a} \new{a computing platform for solving partial differential equations. The charge density $n$ is the integral over the 2D density of states $-2 e  m^*/\pi \hbar^2$ up to the Fermi energy $E_F$,
    \begin{equation}
        n = -e E_F \frac{2 m^*}{\pi \hbar^2}.
    \end{equation}
    The additional energy cost of increasing $E_F$ reduces the charge density $n$, which is called ``quantum capacitance'' effect.
}
 As dielectric constants we take $\epsilon_{\text{Al}_2\text{O}_3} = 9$ for Al$_2$O$_3$, $\epsilon_{\text{hBN}} = 3.8$ for hBN~\cite{Laturia2018} and $\epsilon_{\text{BLG}} = 3.2$ for BLG.
 With $V_{\mathrm{FG}}=8$~V, $V_{\mathrm{BG}}=-4.5~V$ and an additional top gate voltage 
 of $V_{\mathrm{SG}}^\mathrm{eff}=2.2$~V, we obtain the experimentally determined charge carrier density in the channel $n=1.72\times10^{12}$~cm$^{-2}$. 
 \new{The additional top gate accounts for the stray fields caused by the split gates, which are missing in the 2D-cut along the channel, but is also an effective parameter to adjust $n$ in the channel region to the experimental value.}
 Close to the FG, the carrier density decreases and a p-n-p junction forms, see Fig.~3(b).

We approximate the density of states of graphene in the Landau level regime with magnetic field $B$ as
    \begin{equation}
    \rho(E) = \sum_\nu \, \frac{4 eB}{2 \pi h} \frac{\Gamma}{\left( E - E_\nu\right)^2 +  \left( \Gamma/2 \right)^2},
    \end{equation}
with Landau level energy $E_\nu = \hbar \omega_c \sqrt{\nu(\nu-1)}$, \new{and cyclotron frequency} $\omega_c = eB/m^*$~\cite{UnconventionalQHE}, where $\nu$ is the Landau level index. We assumed a broadening of Landau states of $\Gamma = 1$~meV. The charge carrier density is then given by the integration up to the (fixed) Fermi energy
\begin{equation}
    n=\int_0^{E_{\mathrm{F}}} \rho(E)\, \mathrm d E.
\end{equation}
 
 In our approximation, this amounts to a modulation in $n$ by up to $\pm 2\%$ at $B=2$~T.

When this $n(B)$ is inserted back as the charge density in the channel region it electrostatically shifts the dot potential. This shift is entirely classical, and determined by the charge density modulation in the channel, the exact geometry of the device and the dielectric constants. Since $n(B)$ oscillates with $1/B$, the potential under the finger gate also oscillates. Fig.~3(d) shows the calculated electrostatically induced shift of the potential or the QD energy, $\Delta E$ as function of $1/B$. We find good qualitative agreement with the experimental data (c.f. Fig.~3(c)) despite the simplistic 2D model. The oscillation frequencies in Fig.~3(c) and (d) are both determined by $n$. A lower value for the amplitude of $\Delta E$ is expected in the experiment compared to the simulations, as the potential profile in the channel varies due to stray fields from the SG and also the FG. The SdH oscillations in all regions with non-constant potential average out. Only the central region of the channel with equal potential collectively shifts the dot potential. This is overestimated by our 2D model calculation, which explains the larger oscillation amplitude of $\Delta E$ compared to the experimental data. Furthermore, we keep the Fermi energy fixed in the simulation. A slightly oscillating Fermi energy would have a counteracting effect on the amount of charge in the channel and thus lead to less amplitude in $\Delta E$.

 \begin{figure}[t!]
\centering
\includegraphics[draft=false,keepaspectratio=true,clip,width=1.0\linewidth]{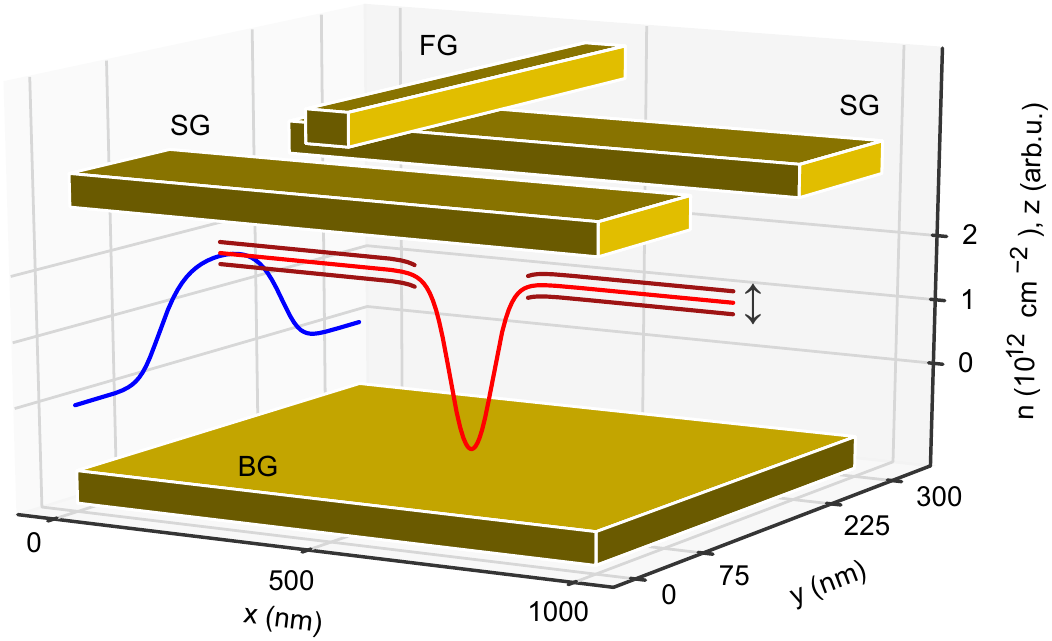}
\caption[Fig04]{
Self-consistent solution of the Poisson equation along a $x$-$z$ and a $y$-$z$ two-dimensional cut in the presence of the experimental gates and an additional top gate (not shown). 
The amplitude of the carrier density fluctuations caused by the SdH-oscillations in a magnetic field according to Eq.~\ref{e:Oscillation} are indicated (dark red lines, not to scale). 
} 
\label{f2}
\end{figure}
 
 As typical QD experiments focus on the energy shift of quantum mechanical levels of a QD in a perpendicular magnetic field, SdH oscillations are seen as an undesired perturbation. In order to suppress such oscillations, we suggest to reduce the channel width, shifting the onset of SdH oscillations to higher fields or to modulate the carrier density along the channel by a few percent using multiple finger gates in order to average out the SdH oscillations.  

In summary, we have shown that a gate-defined BLG QD can be used as a very sensitive \new{probe} for its electrostatic environment. We are able to indirectly probe the density of states in the neighboring channel region connecting the QD to the source and drain reservoirs as a function of a perpendicular magnetic field. This method allowed resolving SdH oscillations from which we extracted the carrier density in the lead region, as well as the amplitude of the oscillations in the QD potential. The experimental data are in good agreement with the results from electrostatic simulations.

\textbf{Acknowledgements} This project has received funding from the European Union's Horizon 2020 research and innovation programme under grant agreement No. 785219 (Graphene Flagship) and from the European Research Council (ERC) under grant agreement No. 820254, the Deutsche Forschungsgemeinschaft (DFG, German Research Foundation) under Germany's Excellence Strategy - Cluster of Excellence Matter and Light for Quantum Computing (ML4Q) EXC 2004/1 - 390534769, through DFG (STA 1146/11-1), and by the Helmholtz Nano Facility~\cite{Albrecht2017May}.  T.F.~and F.L.~acknowledge support by FWF project I-3827 and WWTF project MA14-002. Growth of hexagonal boron nitride crystals was supported by the
Elemental Strategy Initiative conducted by the MEXT, Japan ,Grant Number
JPMXP0112101001,  JSPS KAKENHI Grant Numbers JP20H00354 and the
CREST(JPMJCR15F3), JST.

%
%

\bibliography{literature}

\end{document}